\documentstyle[aasms4,epsfig]{article}

\newcommand\beq{\begin{equation}}
\newcommand\eeq{\end{equation}}

\received{}
\accepted{}      

\begin{document}
{\title{Multicolor photometry of 145 of the HII regions in M33}  

\author{
Linhua Jiang\altaffilmark{1,2},
Jun Ma\altaffilmark{1},
Xu Zhou\altaffilmark{1},
Jiansheng Chen\altaffilmark{1},
Hong Wu\altaffilmark{1}, 
Zhaoji Jiang\altaffilmark{1}, 
Suijian Xue\altaffilmark{1},
and
Jin Zhu\altaffilmark{1} 
}

\altaffiltext{1}
{National Astronomical Observatories of CAS and
CAS-Peking University joint Beijing Astronomical Center,
Beijing, 100012, P. R. China }   

\altaffiltext{2}
{Department of Astronomy,
Peking University, Beijing, 100871, P. R. China }        

\begin{abstract}

This paper is the first in a series presenting CCD multicolor
photometry for 145 HII regions, selected from 369 candidate 
regions from Boulesteix et al. (1974), in the nearby spiral galaxy 
M33. The observations, which covered the whole area of M33, were 
carried out by the Beijing Astronomical Observatory $60/90$ cm 
Schmidt Telescope, in 13 intermediate-band filters, 
covering a range of wavelength from 3800 to 10000{\AA}. 
This provides a series of maps which can be converted to 
a multicolor map of M33, in pixels 
of $1\arcsec{\mbox{}\hspace{-0.15cm}.} 7
\times 1\arcsec{\mbox{}\hspace{-0.15cm}.} 7$.
Using aperture photometry we obtain the spectral energy distributions 
(SEDs) for these HII regions. We also give their identification charts.
Using the relationship between the BATC intermediate-band system used 
for the observations and the {\it UBVRI} broad-band system, the magnitudes 
in the {\it B} and {\it V} bands are then derived. Histograms of the 
magnitudes in {\it V} and in {\it B$-$V} are plotted, and the 
color-magnitude diagram is also given. The distribution of magnitudes 
in the {\it V} band shows that the apparent magnitude of almost all the 
regions is brighter than 18, corresponding to an absolute magnitude of 
$-$6.62 for an assumed distance modulus of 24.62, which corresponds to 
a single main sequence O5 star, while the distribution of color 
shows that the sample is blue, with a mode close to $-$0.05 
as would be expected from a range of typical young clusters.    

\end{abstract}
   

\section{INTRODUCTION} 

HII regions provide an excellent means of studying the ongoing and 
accumulated star formation in a late type galaxy. In a well-resolved 
galaxy, HII regions can offer much useful information: about the 
current status of star formation, and the physical conditions in the  
interstellar medium near the hot young stars within the galaxy 
(see e.g., Wyder, Hodge, \& Skelton 1997), and it is well established 
that observations of extragalactic HII regions are important for the 
understanding of the chemical evolution and star formation history of 
the parent galaxy. For these purposes HII regions in many nearby 
galaxies have been explored in considerable detail 
(Garnett \& Shields 1987), which include M31 
(Baade \& Arp 1964; Pellet et al. 1978; Walterbos \& Braun 1992; 
Walterbos 2000), and M81 (Garnett \& Shields 1987; Petit, Sivan, \& 
Karachentsev 1988; Hill et al. 1995; Miller \& Hodge 1996).

M33, one of our two nearest spiral neighbors, is a small Scd Local 
Group galaxy, whose distance modulus, determined via its Cepheids 
by Freedman,Wilson and Madore (1991), was more recently revised to 
24.62 by Freedman et al. (2001). Because it is so near, and has quite 
a low inclination ($\sim 33^\circ$), M33 is an excellent candidate 
for HII region studies. A good database for its star clusters has 
been built up from the ground-based work (Hiltner 1960; Kron \& 
Mayall 1960; Melnick \& D'Odorico 1978; Christian \& Schommer 1982, 
1988}), and from the {\it {Hubble Space Telescope (HST)}} images 
(Chandar, Bianchi, \& Ford 1999a, 1999b; \cite{Chandar01}).
Ma et al. (2001, 2002a, 2002b) obtained the spectral energy 
distributions (SEDs) of 144 star clusters, and estimated their ages 
by comparing the photometry of each object with theoretical stellar 
population synthesis models for different values of metallicity.

The HII regions in M33 have proved of interest to astronomers for
well over half a century (see e.g., Aller 1942, Haro 1950, Court\`{e}s 
\& Cruvellier 1965).  Aller (1942) used the ratio of
[OIII] $\lambda5007$/[OII] $\lambda3727$ to be the criterion for
identifying HII regions. However, most studies searched HII regions
by photographic $H\alpha$ survey.
In 1974, Boulesteix et al. (1974) compiled a major catalogue 
of 369 distinct HII regions by a complete photographic $H\alpha$ survey of M33. 
In this study, Boulesteix et al. (1974) defined an HII region
within the  $H\alpha$ image of M33 to correspond to an emission
measure of $\rm {150~cm^{-6}pc}$, i.e. three times above the
rms noise level.
Later Court\`{e}s et al. (1987) added an additional 410 objects to 
that catalogue, also using a photographic $H\alpha$ survey, and
defined an HII region
within the  $H\alpha$ image of M33 to correspond to an emission
measure of $\rm {40~cm^{-6}pc}$.
In a paper in which they selected isolated candidates for
SS433-like stellar systems, Calzetti et al. (1995) reported 432 
compact regions emitting $H\alpha$. Recently a more extensive 
catalogue with 2338 $H\alpha$ emission regions of M33 was published 
by Hodge et al. (1999). While these authors made an explicit position 
and luminosity catalogue, others paid more attention to the luminosity 
function and size distribution of the HII regions on the basis of
$H\alpha$ surveys (e.g., Wyder et al. 1997, Cardwell et al. 2000).
The latter authors claim to have identified over 10000 separate 
HII regions in $H\alpha$, but so far have not published a catalogue.          

Although there is already much observational information in the
literature, for full physical information about the HII regions 
it is of great value to obtain multiband photometry, which can 
provide accurate SEDs. In the present study we present CCD 
spectrophotometry of a set of HII regions in M33 using images 
obtained with the Beijing-Arizona-Taiwan-Connecticut (BATC) 
Multicolor Sky Survey Telescope, designed to obtain SED information
for galaxies (Fan et al. 1996). The BATC system uses the 60/90 cm 
Schmidt telescope with 15 intermediate bandwidth filters. 
Here we use 13 of these filters, from 3800 to 10000{\AA}, with images 
covering the whole visible extent of M33. In this paper we present 
the SEDs of 145 HII regions selected from the sample of 
Boulesteix et al. (1974), and also supply identification charts
for these regions. We go on to compute the {\it B} and {\it V} magnitudes 
for these regions, using previously derived relationships between 
the BATC intermediate band system and the {\it UBVRI} broad-band system. 
Finally we plot histograms of the {\it V} band magnitudes and also of 
{\it B$-$V}, as well as plotting the corresponding color-magnitude 
diagram for our sample regions.     

We note that the extinction for almost all of the HII regions
remains undetermined. Although it has been derived for a few of 
the large regions (Viallefond, Donas, \& Goss 1983; Viallefond \& 
Goss 1986; Churchwell \& Goss 1999), the extinction for the regions 
sampled here has not been previously presented. In M81 the extinction 
varies relatively little from HII region to HII region
(Hill et al. 1995), but in M33 there is considerable variation 
(Viallefond, Donas \& Goss 1983). We cannot use the assumption of 
uniform extinction to correct for reddening in the present paper, 
so we have not been able to infer the intrinsic fluxes of the regions. 
In a forthcoming article (Jiang et al. 2002) we will use the flux ratio
of $H_{\alpha}/H_{\beta}$ to study the extinction region by region. 

The outline of this paper is as follows: details of the observations
and data reduction are given in section 2. In section 3 we give the 
SEDs and the identification charts for the 145 regions studied. 
In this section also we produce the histograms of the {\it V} magnitude, 
and of {\it B$-$V}, as well as the color-magnitude diagram. 
Finally in section 4 we give a summary.

\section{SAMPLE OF HII REGIONS, OBSERVATIONS AND DATA REDUCTION}

\subsection{Selection of the Sample }

The sample of HII regions chosen for this paper comes from the
catalogue of 369 HII regions by Boulesteix et al. (1974). 
Court\`{e}s et al. (1987) added an additional 410 objects to this 
catalogue. We take regions from Boulesteix's catalogue as our sample, 
while we use positions and sizes available from Court\`{e}s et al. (1987), 
because positions presented by Court\`{e}s et al. (1987) are more
accurate than those given by Boulesteix et al. (1974) (spatial resolutions 
are about $1\arcsec$ and $4\arcsec$ respectively). We set out to 
observe all the Boulesteix regions, but found that for 224 of these 
the signal-to-noise ratio achieved was not sufficient in at least 
some of the filters, so we have eliminated these from the final sample 
presented. For this purpose we used the criterion that a region with
a photometric uncertainty in any filter of more than 1 magnitude was 
not suitable for analysis. This criterion rejects the least luminous
regions, which makes our sample regions somewhat bright (see section 3 
and Fig. 1). At last the 145 HII regions which passed this limit 
test are included here.  

\subsection{BATC Observations}

As M33 is so near, it subtends a large angle on the sky, which makes
it non-trivial to obtain CCD images over the whole disk (Walterbos 2000).
This difficulty can be resolved with an instrument with a large field format.
The BATC telescope and filter system is well suited to this type of
observations. The telescope is a $60/90$ cm f/3 Schmidt, belonging to the
Beijing Astronomical Observatory (BAO) at the Xinglong station. A Ford
Aerospace 2048$\times$2048 CCD camera, with 15 $\mu$m pixel size is 
mounted at the Schmidt focus, giving a field of view of 
$58^{\prime}$ $\times $ $58^{\prime}$ with a pixel size of
$1\arcsec{\mbox{}\hspace{-0.15cm}.} 7$. 

The multiband BATC filter system comprises 15 intermediate-band
filters, covering the full optical wavelength range from 3000 to 
10000{\AA}. The filters were designed specifically to avoid 
contamination from the brightest and most variable night sky emission 
lines. Details of the BAO Schmidt Telescope, the CCD camera, the data 
acquisition system, and the definition of the BATC filter passbands 
can be found in previous publications (Fan et al. 1996; Zheng et al. 
1999). The observations of M33 were carried out from September 23, 1995, 
through August 28, 2000, with a total exposure time of about 38 hr 
15 min. The CCD images, which were centered at 
${\rm RA=01^h33^m50^s{\mbox{}\hspace{-0.13cm}.}58}$ and
DEC=30$^\circ39^{\prime}08^{\prime\prime}{\mbox{}\hspace{-0.15cm}.4}$  
(J2000) covered the full optical extent of M33, imaging this in 13 of 
the BATC filters. The dome flats were obtained using
a diffusor plate in front of the Schmidt corrector plate. For flux 
calibration the Oke-Gunn primary standard stars HD~19445, HD~84937, 
BD~+26$^\circ$2606, and BD~+17$^\circ$4708 were observed under 
photometric conditions (see Yan et al. 2000, Zhou et al. 2001 for 
details). The parameters of the filters, and the basic statistics 
of the observations are given in Table 1.   

\setcounter{table}{0} 
\begin{table}[ht]
\caption[]{Parameters of the BATC filters
and statistics of observations}
\vspace {0.5cm}
\begin{tabular}{cccccc}
\hline
\hline
 No. & Name& cw\tablenotemark{a}~~(\AA)& Exp. (hr)&  N.img\tablenotemark{b}
 & rms\tablenotemark{c} \\
\hline
1  & BATC03& 4210   & 00:55& 04 &0.024\\
2  & BATC04& 4546   & 01:05& 04 &0.023\\
3  & BATC05& 4872   & 03:55& 19 &0.017\\
4  & BATC06& 5250   & 03:19& 15 &0.006\\
5  & BATC07& 5785   & 04:38& 17 &0.011\\
6  & BATC08& 6075   & 01:26& 08 &0.016\\
7  & BATC09& 6710   & 01:09& 08 &0.006\\
8  & BATC10& 7010   & 01:41& 08 &0.005\\
9  & BATC11& 7530   & 02:07& 10 &0.017\\
10 & BATC12& 8000   & 03:00& 11 &0.003\\
11 & BATC13& 8510   & 03:15& 11 &0.005\\
12 & BATC14& 9170   & 05:45& 25 &0.011\\
13 & BATC15& 9720   & 06:00& 26 &0.009\\      
\hline
\end{tabular}\\
\tablenotetext{a}{Central wavelength for each BATC filter}
\tablenotetext{b}{Image numbers for each BATC filter}
\tablenotetext{c}{Zero point error, in magnitude, for each filter
as obtained from the standard stars}
\end{table}

\subsection{Data Reduction}

Using standard procedures, which include bias subtraction and 
flat-fielding of the CCD images, the data were reduced by automatic 
reduction software: PIPELINE I, developed for the BATC multicolor 
sky survey (Fan et al. 1996; Zheng et al. 1999). Flat-fielded images 
for a given band were combined by integer pixel shifting. The images 
were re-centered and position calibrated, basing the astrometry on 
stars from the {\it HST} Guide Star Catalogue. The technique of integer 
pixel shifting is somewhat rather crude for registering images, since a pixel
corresponds to $1\arcsec{\mbox{}\hspace{-0.15cm}.} 7$. However, as M33 
is quite close, $1\arcsec{\mbox{}\hspace{-0.15cm}.} 7$ does not
amount to a major fraction of the radius of a large HII region.
Cosmic rays and bad pixels were 
corrected by comparison between images during combination. The images 
were calibrated using observations of standard stars, having convolved 
the SEDs of these stars with the measured BATC filter transmission 
functions (Fan et al. 1996), to derive the fluxes of these Oke-Gunn 
standards as observed through the BATC filters. In Table 1, {\it Column 6}  
gives the zero point errors in magnitude for the standard stars
through each filter. The formal errors obtained for these stars in 
the 13 BATC filters used are $\la 0.02$ mag, which implies that we can 
define photometrically the BATC system to an accuracy of better than 
0.02 mag.    

\section{RESULTS}

\subsection{Integrated Photometry}

To obtain the magnitude of a given HII region, we used aperture 
photometry. The HII regions have a variety of shapes: disks, loops, arcs, 
and filaments, and a considerable range of sizes (see Court\`{e}s 
et al. 1987). We followed the specifications of Court\`{e}s et al. 
who presented measured sizes in two orthogonal coordinates: the E-W 
extent, and the N-S extent, for all regions; we used the larger of these 
as the aperture for our photometric measurements. As the HII regions 
are found in a variety of local environments, background subtraction 
is not straightforward. We determined the background within an annular 
aperture for each region. Within the chosen annulus we fitted each pixel 
row of the image by a linear median, obtaining a surface, and repeated 
this process in the column direction with this surface. After this, we 
rejected points which differed from the mean background by over $30\%$, 
and derived a smoothed background fit as the final step. Using this
background surface, we made the background subtraction for each HII region. 
We used a standard set of ratios between the aperture radius for a given
region, and the inner and outer radii of its background annuli; these ratios
are specified in Table 2, where {\it Column 1} lists a set of aperture 
ranges in pixels, while {\it Column 2} and {\it 3} give the corresponding 
inner and outer radii for the background apertures. For example, 
if an HII region is 30 pixels in size (diameter), i.e. photometric aperture 
radius is 15, which is more than 10 and less than 20, then the inner radius 
of its background annulus is $15+15=30$, and the outer radius, $15+25=40$.
Using these parameters and the method outlined, the backgrounds were 
obtained and subtracted for the complete set of HII regions in each filter.

\setcounter{table}{1} 
\begin{table}[ht]
\caption[]{The relationships between the range of the photometric aperture size
and the inner and outer sizes of its background annulus}
\vspace {0.5cm}
\begin{tabular}{ccc}
\hline
\hline
 Radius of photometric aperture (r) & Inner radius of annulus & Outer radius of annulus
\\
\hline
{$\leq 5$}     & r+5     & r+15\\
{$5\sim 10$}  & r+10    & r+20\\
{$10\sim20$}  & r+15    & r+25\\
{$20\sim30$}  & r+20    & r+30\\
{$30\sim40$}  & r+25    & r+35\\
{$40\sim50$}  & r+30    & r+40\\
{$\geq 50$}    & r+40    & r+50\\
\hline
\end{tabular}\\
\end{table}

Finally we used the fluxes in the filters to derive the SEDs for the
145 HII regions, which are listed in Table 3. This contains the following
information: {\it Column 1} is a check number for each HII region, taken 
from Boulesteix et al. (1974), {\it Column 2} through {\it 14} show the 
magnitudes in the selected BATC bands, and on a second row for each HII 
region in these columns we give the magnitude uncertainty for each band. 
These uncertainties include the errors from the object count rate, the 
sky variance, and the instrument gain.       

\subsection{Magnitudes in the B and V Bands, and (B$-$V) Colors}
 
Using Landolt standards, Zhou et al. (2002) derived the relationships
between the BATC intermediate band system and the {\it UBVRI} broad-band system, 
making use of the catalogues of Laondolt (1983, 1992), and of 
Galad\'\i-Enr\'\i quez et al. (2000). The relationships are given
in equations (1) and (2) as :
\beq
m_B=m_{04}+(0.2218\pm0.033)(m_{03}-m_{05})+0.0741\pm0.033,
\eeq
\beq
m_V=m_{07}+(0.3233\pm0.019)(m_{06}-m_{08})+0.0590\pm0.010.
\eeq
Using equations (1) and (2) we transformed the magnitudes of the 145
HII regions in the BATC03, BATC04 and BATC05 bands into {\it B} band magnitudes, 
and also derived {\it V} band magnitudes from those in BATC06, BATC07 and BATC08. 
HII regions in general have strong emission lines, and in the bands chosen
here, the BAT05 band can be significantly affected. In those cases where 
there is a strong emission line in the BATC05 band, it is suited to 
interpolate between the BATC04 and BATC06 bands to substitute the BATC05 band. Here 
we take the mean value between BATC04 and BATC06 as the value of BATC05 to calculate 
the {\it B} magnitude, when there is strong emission line in the BATC05 band. 

In Figure 1 we give a histogram of the distribution of magnitudes
in the computed {\it V} band. As is mentioned previously, our criterion 
of choosing the sample rejects the least luminous HII regions (see section 
2.1 for a detail), and it makes our sample 
regions somewhat bright. From this figure we can see that almost all of 
our sample of HII regions are brighter than 18th magnitude, and using 
a distance modulus for the galaxy of 24.62 (Freedman et al. 2001) this 
corresponds to an absolute magnitude in {\it V} of $-$6.62. This is the 
equivalent of a singe main sequence O5 star, and although the stars 
within different HII regions may differ in their mass distribution, 
this gives us an idea of the observational limits of our data set. 
In fact, it is generally accepted that the bright HII regions are ionized
by several to hundreds of young massive stars (see e.g., Kennicutt 1988; 
Kennicutt \& Chu 1988; Mayya 1994; Oey \& Kennicutt 1997), especially O 
stars, and O5 stars may be the most typical (see e.g., Kennicutt 1984; 
Kennicutt 1988; Kennicutt \& Chu 1988; Mayya 1994), whose magnitudes 
are exactly the lower magnitude limit of our sample regions. For example, 
taking the bright HII regions in nearby galaxies as his research sample, 
Kennicutt (1984, 1998) gave the equivalent number of O5 {\it V} stars
which would be required 
to ionize the sample regions, and found that the number ranged from a few 
to about a thousand. Here we take the 
brightest HII region in M33, NGC604 (No. 608 in this text), as a sample 
to compute the equivalent number of O5 stars which would ionize the 
region. Using equation (2) we get {\it V} magnitude of NGC604, 12.40, which 
corresponds to an absolute {\it V} magnitude of $-$12.22. By comparing this 
value with {\it V} magnitude of a single main sequence O5 star, we get the 
equivalent number of O5 stars in NGC604, 174. This result is well in 
agreement with one given by Hunter et al. (1996), 186 O stars,
and also in the range between 150 and 215 derived by 
Gonz\'{a}lez Delgado (2000).

We present the distribution of the HII regions in our computed {\it B$-$V} 
colors in Figure 2. We can see that in general the regions are quite blue,
i.e. the colors of $90\%$ sample HII regions are bluer than 0.4,
with a mode close to $-$0.05 as
expected for zones whose light is dominated by stars in young clusters.
In fact, Kennicutt et al. (1988) defined the blue populous
clusters in M33 to be those with {\it B$-$V}$\la 0.5$. If so,
most of our sample regions are dominated by young blue clusters,
as is already expected. By photometry in {\it BVR} continuum bands
and in the emission line of $H\alpha$+[NII], Mayya (1994) derived
{\it B$-$V} colors for 186 HII regions in 9 galaxies, and found
that the colors of the most sample regions are between $-0.2$
and 0.5, and the median value is 0.21. It is obvious that
our sample HII regions are typical as those of Mayya (1994).
The color-magnitude diagram, {\it V} 
against {\it B$-$V}, is shown in Figure 3.

\subsection{The SEDs and the HII region identification charts}

The SEDs for each of the HII regions observed are plotted in Figure 4.
For convenience, we calculated the ratio of the flux in each filter 
to the flux in filter BATC08, and these ratios are shown as values
of y-axis in Figure 4. From this figure,  
we can see that many of the HII regions have strong emission lines, 
which show up in the BATC05, BATC09 and BATC14 bands. These lines 
clearly correspond to [OIII] at 5007{\AA}, the combination of $H\alpha$ 
at 6563A and the [NII] doublet surrounding it, and the [SIII] line at 
9069{\AA} respectively. 

In Figure 5, we give identification charts for all of the HII 
regions observed in the BATC07 band, using positions from 
Court\`{e}s et al. (1987). The diameter of the circle gives the diameter 
of the photometric aperture 
used in each case. Where a region is not included in the catalogue of 
Court\`{e}s et al. (1987) we used positions and effective radii from 
Boulesteix et al. (1974). For clarity, in Fig. 6 we have divided the 
image of M33 from Fig. 5 into ten subfields, and have provided an amplified 
version of the finding chart in the 10 separate subfields used.    

\section{SUMMARY}

Using images of M33 obtained with the Beijing-Arizona-Taiwan-Connecticut
(BATC) multi-band Sky Survey Telescope, we present CCD spectrophotometry 
of HII regions in this galaxy. The main result is the spectral energy 
distributions (SEDs) of 145 HII regions, calalogued by Boulesteix et al. 
(1974), for which we also provide identification charts. We have been able 
to use the information in the 13 intermediate bands observed to transform 
to {\it V} magnitude and {\it B$-$V} color, using previously derived 
relationships between the BATC system and the {\it UBVRI} system. We have 
plotted histograms of the distribution of the {\it V} magnitudes and the 
{\it B$-$V} colors, as well as a {\it V} versus {\it B$-$V} 
color-magnitude diagram for the 145 regions studied. 
Information on the observational limitation of the present survey can be 
inferred from the fact that virtually all the regions studied have 
apparent magnitude 18 or brighter, and from the {\it B$-$V} histogram we 
can infer that the regions observed are generally quite blue, as expected 
for zones whose dominant light sources are young clusters.   
                                                           
\acknowledgements
We would like to thank the anonymous referee for his/her
insightful comments and suggestions that improved this paper.
We are also indebted to him for English editing.
The work is supported partly by the National Sciences
Foundation under the contract No. 19833020 and No. 19503003. 
The BATC Survey is supported by the
Chinese Academy of Sciences, the Chinese National Natural Science
Foundation and the Chinese State Committee of Sciences and
Technology. 
The project is also supported in part
by the National Science Foundation (grant INT 93-01805) and
by Arizona State University, the University of Arizona and Western
Connecticut State University.

\clearpage
{\small
\setcounter{table}{2}
\begin{table}[ht]
\caption{SEDs of 145 HII Regions in M33}
\vspace {0.3cm}
\begin{tabular}{cccccccccccccc}
\hline
\hline
 No. & 03  &  04 &  05 &  06 &  07 &  08 &  09 &  10 &  11 &  12 &  13 &  14 &  15\\
(1)    & (2) & (3) & (4) & (5) & (6) & (7) & (8) & (9) & (10) & (11) & (12) & (13) & (14)\\
\hline
     1 &  16.87 &  16.88 &  16.61 &  16.91 &  16.88 &  17.00 &  16.29 &  17.04 &  17.09 &  17.04 &  17.12 &  16.95 & 17.38 \\        &   0.09 &   0.10 &   0.08 &   0.11 &   0.13 &   0.15 &   0.08
&   0.15 &   0.16 &   0.24 &   0.36 &   0.21 &  0.43 \\
     4 &  17.36 &  17.66 &  15.94 &  18.08 &  17.79 &  18.24 &  14.90 &  17.10 &  16.73 &  16.43 &  16.37 &  15.25 & 15.74 \\        &   0.23 &   0.31 &   0.07 &   0.51 &   0.48 &   0.73 &   0.04
&   0.26 &   0.19 &   0.22 &   0.30 &   0.07 &  0.16 \\
     5 &  15.98 &  15.94 &  15.55 &  16.01 &  16.06 &  16.18 &  15.56 &  16.31 &  16.42 &  16.38 &  16.51 &  16.26 & 16.45 \\        &   0.05 &   0.05 &   0.04 &   0.06 &   0.07 &   0.08 &   0.05
&   0.09 &   0.10 &   0.14 &   0.22 &   0.12 &  0.20 \\
     6 &  16.75 &  16.63 &  16.69 &  16.79 &  16.97 &  17.03 &  17.77 &  16.94 &  16.74 &  17.10 &  16.87 &  16.68 & 17.05 \\        &   0.12 &   0.11 &   0.13 &   0.14 &   0.20 &   0.22 &   0.43
&   0.20 &   0.17 &   0.36 &   0.42 &   0.23 &  0.46 \\   
    7A &  15.54 &  15.59 &  15.46 &  15.70 &  15.90 &  16.04 &  15.54 &  15.99 &  15.82 &  15.93 &  16.15 &  15.75 & 15.63 \\        &   0.04 &   0.05 &   0.04 &   0.05 &   0.08 &   0.09 &   0.06
&   0.08 &   0.07 &   0.12 &   0.21 &   0.10 &  0.12 \\
    7B &  15.76 &  15.70 &  15.64 &  15.64 &  15.59 &  15.60 &  15.47 &  15.33 &  15.16 &  15.15 &  15.03 &  14.83 & 14.88 \\        &   0.05 &   0.05 &   0.05 &   0.05 &   0.05 &   0.05 &   0.05
&   0.04 &   0.04 &   0.05 &   0.07 &   0.04 &  0.06 \\
    8A &  14.52 &  14.52 &  14.47 &  14.64 &  14.75 &  14.85 &  14.54 &  14.88 &  14.93 &  14.97 &  15.01 &  14.78 & 14.76 \\        &   0.02 &   0.02 &   0.03 &   0.03 &   0.04 &   0.04 &   0.03
&   0.04 &   0.05 &   0.07 &   0.10 &   0.06 &  0.08 \\
     9 &  16.26 &  16.21 &  16.03 &  16.17 &  16.17 &  16.24 &  16.06 &  16.17 &  16.02 &  16.14 &  16.21 &  15.84 & 15.99 \\        &   0.05 &   0.05 &   0.04 &   0.05 &   0.06 &   0.06 &   0.05
&   0.06 &   0.06 &   0.08 &   0.11 &   0.06 &  0.09 \\
    11 &  16.72 &  16.64 &  16.37 &  16.58 &  16.64 &  16.78 &  16.15 &  16.77 &  16.73 &  16.81 &  16.90 &  16.61 & 16.86 \\        &   0.10 &   0.10 &   0.09 &   0.10 &   0.14 &   0.16 &   0.09
&   0.15 &   0.15 &   0.25 &   0.38 &   0.19 &  0.34 \\
    12 &  17.63 &  17.56 &  17.43 &  17.60 &  17.58 &  17.64 &  17.16 &  17.63 &  17.76 &  17.76 &  17.90 &  17.83 & 17.57 \\        &   0.16 &   0.16 &   0.16 &   0.19 &   0.23 &   0.24 &   0.16
&   0.24 &   0.28 &   0.43 &   0.69 &   0.43 &  0.48 \\
    13 &  16.08 &  16.10 &  15.37 &  16.20 &  16.09 &  16.18 &  14.60 &  16.05 &  16.15 &  16.07 &  16.18 &  15.51 & 16.15 \\        &   0.07 &   0.07 &   0.04 &   0.09 &   0.10 &   0.11 &   0.03
&   0.09 &   0.10 &   0.15 &   0.24 &   0.09 &  0.22 \\
    17 &  15.80 &  15.81 &  15.43 &  15.84 &  15.83 &  15.92 &  14.81 &  15.90 &  15.87 &  15.77 &  15.76 &  15.40 & 15.77 \\        &   0.06 &   0.06 &   0.04 &   0.06 &   0.08 &   0.09 &   0.03
&   0.08 &   0.08 &   0.12 &   0.17 &   0.08 &  0.16 \\  
    21 &  16.30 &  16.39 &  16.18 &  16.32 &  16.43 &  16.51 &  16.05 &  16.51 &  16.34 &  16.33 &  16.42 &  16.34 & 16.06 \\        &   0.10 &   0.11 &   0.10 &   0.11 &   0.15 &   0.17 &   0.11
&   0.16 &   0.14 &   0.22 &   0.36 &   0.21 &  0.24 \\
    25 &  14.46 &  14.15 &  13.93 &  13.91 &  13.72 &  13.70 &  13.51 &  13.59 &  13.60 &  13.53 &  13.50 &  13.43 & 13.47 \\        &   0.02 &   0.02 &   0.02 &   0.02 &   0.01 &   0.01 &   0.01
&   0.01 &   0.01 &   0.02 &   0.02 &   0.02 &  0.02 \\
    26 &  16.51 &  16.52 &  16.45 &  16.71 &  16.73 &  16.76 &  16.26 &  16.90 &  17.07 &  17.15 &  17.23 &  17.28 & 17.43 \\        &   0.07 &   0.07 &   0.07 &   0.09 &   0.12 &   0.12 &   0.08
&   0.14 &   0.16 &   0.27 &   0.41 &   0.29 &  0.46 \\
    34 &  17.91 &  17.77 &  17.29 &  17.87 &  17.58 &  17.74 &  16.87 &  17.60 &  17.50 &  17.46 &  17.65 &  17.10 & 17.14 \\        &   0.22 &   0.20 &   0.14 &   0.24 &   0.23 &   0.27 &   0.12
&   0.23 &   0.22 &   0.33 &   0.57 &   0.22 &  0.33 \\
    38 &  16.04 &  16.04 &  15.79 &  16.36 &  16.42 &  16.72 &  15.45 &  16.74 &  16.64 &  16.59 &  16.66 &  15.91 & 16.24 \\        &   0.09 &   0.10 &   0.09 &   0.15 &   0.20 &   0.26 &   0.08
&   0.27 &   0.25 &   0.35 &   0.52 &   0.18 &  0.34 \\
    40 &  15.51 &  15.51 &  15.32 &  15.60 &  15.85 &  16.08 &  14.70 &  16.05 &  16.18 &  16.16 &  16.31 &  15.67 & 16.27 \\        &   0.05 &   0.06 &   0.05 &   0.07 &   0.10 &   0.13 &   0.04
&   0.13 &   0.15 &   0.21 &
  0.34 &   0.13 &  0.31 \\
\end{tabular}
\end{table}                       
}                                          

\clearpage
{\small  
\setcounter{table}{2}
\begin{table}[ht]
\caption{Continued}
\vspace {0.3cm}
\begin{tabular}{cccccccccccccc}
\hline
\hline
 No. & 03  &  04 &  05 &  06 &  07 &  08 &  09 &  10 &  11 &  12 &  13 &
14 &  15\\
(1)    & (2) & (3) & (4) & (5) & (6) & (7) & (8) & (9) & (10) & (11) & (12) & (13) & (14)\\
\hline
    49 &  14.06 &  14.16 &  12.96 &  14.21 &  14.43 &  14.94 &  12.15 &  14.38 &  14.56 &  14.87 &  15.46 &  13.35 & 15.46 \\        &   0.04 &   0.05 &   0.02 &   0.05 &   0.09 &   0.14 &   0.01
&   0.08 &   0.10 &   0.21 &   0.54 &   0.05 &  0.49 \\
    55 &  17.89 &  17.65 &  17.05 &  17.26 &  16.86 &  16.92 &  16.32 &  16.67 &  16.67 &  16.48 &  16.64 &  16.31 & 16.46 \\        &   0.21 &   0.17 &   0.11 &   0.13 &   0.12 &   0.13 &   0.07
&   0.10 &   0.10 &   0.14 &   0.23 &   0.11 &  0.19 \\
    56 &  16.88 &  16.51 &  16.10 &  16.29 &  16.02 &  16.11 &  15.50 &  15.99 &  16.08 &  16.00 &  15.99 &  15.97 & 16.12 \\        &   0.12 &   0.09 &   0.07 &   0.08 &   0.08 &   0.09 &   0.05
&   0.08 &   0.09 &   0.12 &   0.17 &   0.11 &  0.18 \\
    58 &  17.80 &  17.76 &  17.56 &  17.54 &  17.77 &  17.96 &  17.50 &  18.08 &  18.04 &  18.10 &  18.43 &  18.38 & 18.49 \\        &   0.16 &   0.15 &   0.14 &   0.14 &   0.22 &   0.26 &   0.16
&   0.28 &   0.27 &   0.47 &   0.93 &   0.57 &  0.91 \\
    59 &  16.23 &  16.15 &  15.97 &  16.20 &  16.16 &  16.24 &  15.84 &  16.27 &  16.37 &  16.36 &  16.37 &  16.20 & 16.53 \\        &   0.05 &   0.05 &   0.04 &   0.05 &   0.06 &   0.07 &   0.05
&   0.07 &   0.08 &   0.12 &   0.17 &   0.10 &  0.18 \\ 
    60 &  17.15 &  17.24 &  16.65 &  17.43 &  17.44 &  17.73 &  16.21 &  17.72 &  17.92 &  17.51 &  17.91 &  17.34 & 17.73 \\        &   0.15 &   0.17 &   0.11 &   0.23 &   0.29 &   0.38 &   0.09
&   0.37 &   0.46 &   0.48 &   0.99 &   0.39 &  0.79 \\
    62 &  15.96 &  16.08 &  15.46 &  16.28 &  16.21 &  16.39 &  14.68 &  16.29 &  16.23 &  16.25 &  16.21 &  15.37 & 15.85 \\        &   0.07 &   0.08 &   0.05 &   0.10 &   0.12 &   0.15 &   0.03
&   0.13 &   0.13 &   0.21 &   0.29 &   0.09 &  0.19 \\
    63 &  16.87 &  17.01 &  16.08 &  16.94 &  16.87 &  17.18 &  15.50 &  17.21 &  17.03 &  16.82 &  17.06 &  16.13 & 16.73 \\        &   0.16 &   0.19 &   0.09 &   0.20 &   0.24 &   0.32 &   0.07
&   0.32 &   0.28 &   0.36 &   0.64 &   0.18 &  0.44 \\
    65 &  17.17 &  17.15 &  16.38 &  17.35 &  17.20 &  17.51 &  16.01 &  17.54 &  17.69 &  17.47 &  17.73 &  16.78 & 18.12 \\        &   0.12 &   0.12 &   0.07 &   0.16 &   0.17 &   0.23 &   0.06
&   0.23 &   0.28 &   0.35 &   0.64 &   0.18 &  0.86 \\
    77 &  15.15 &  15.19 &  14.88 &  15.30 &  15.26 &  15.35 &  14.24 &  15.28 &  15.30 &  15.29 &  15.22 &  15.02 & 15.31 \\        &   0.04 &   0.04 &   0.03 &   0.05 &   0.05 &   0.06 &   0.02
&   0.06 &   0.06 &   0.09 &   0.12 &   0.07 &  0.12 \\
    80 &  18.65 &  18.41 &  18.28 &  18.28 &  18.06 &  18.10 &  17.72 &  17.91 &  17.81 &  17.67 &  17.68 &  17.57 & 17.62 \\        &   0.29 &   0.25 &   0.24 &   0.25 &   0.25 &   0.26 &   0.18
&   0.22 &   0.21 &   0.26 &   0.37 &   0.23 &  0.33 \\
    83 &  15.66 &  15.65 &  15.50 &  15.77 &  15.74 &  15.89 &  15.35 &  15.84 &  15.72 &  15.72 &  15.62 &  15.27 & 15.34 \\        &   0.04 &   0.04 &   0.04 &   0.05 &   0.06 &   0.06 &   0.04
&   0.06 &   0.06 &   0.08 &   0.11 &   0.05 &  0.08 \\
    84 &  19.13 &  19.08 &  18.63 &  19.03 &  18.72 &  18.86 &  18.69 &  18.56 &  18.50 &  18.48 &  18.49 &  18.20 & 18.10 \\        &   0.33 &   0.34 &   0.25 &   0.37 &   0.33 &   0.38 &   0.33
&   0.29 &   0.29 &   0.39 &   0.53 &   0.29 &  0.35 \\               
    85 &  16.22 &  16.18 &  16.03 &  16.13 &  16.02 &  16.09 &  15.57 &  15.92 &  15.78 &  15.67 &  15.64 &  15.40 & 15.32 \\        &   0.07 &   0.07 &   0.06 &   0.07 &   0.08 &   0.09 &   0.05
&   0.07 &   0.07 &   0.09 &   0.12 &   0.07 &  0.09 \\
    87 &  15.85 &  15.91 &  15.18 &  15.96 &  15.52 &  15.58 &  14.10 &  15.35 &  15.24 &  15.11 &  15.23 &  14.71 & 14.97 \\        &   0.10 &   0.11 &   0.06 &   0.13 &   0.10 &   0.11 &   0.03
&   0.09 &   0.08 &   0.11 &   0.17 &   0.07 &  0.13 \\
    88 &  14.47 &  14.45 &  13.76 &  14.45 &  14.41 &  14.49 &  13.52 &  14.40 &  14.44 &  14.34 &  14.40 &  14.01 & 14.24 \\        &   0.03 &   0.03 &   0.02 &   0.03 &   0.04 &   0.04 &   0.02
&   0.04 &   0.04 &   0.06 &   0.09 &   0.04 &  0.07 \\
   89B &  16.15 &  16.07 &  16.03 &  16.21 &  16.25 &  16.36 &  16.36 &  16.43 &  16.43 &  16.46 &  16.27 &  16.13 & 16.17 \\        &   0.06 &   0.06 &   0.06 &   0.07 &   0.08 &   0.09 &   0.09
&   0.10 &   0.10 &   0.15 &   0.18 &   0.11 & 0.15 \\
    93 &  14.43 &  14.39 &  14.01 &  14.26 &  14.41 &  14.36 &  13.62 &  14.23 &
 14.17 &  14.83 &  15.44 &  14.39 & 15.18 \\        &   0.02 &   0.03 &   0.02 &
  0.03 &   0.03 &   0.03 &   0.02 &   0.03 &   0.03 &   0.07 &   0.16 &   0.05 &
 0.12 \\
\end{tabular}
\end{table}          
}

\clearpage
{\small 
\setcounter{table}{2}
\begin{table}[ht]
\caption{Continued}
\vspace {0.3cm}
\begin{tabular}{cccccccccccccc}
\hline
\hline
 No. & 03  &  04 &  05 &  06 &  07 &  08 &  09 &  10 &  11 &  12 &  13 &
14 &  15\\
(1)    & (2) & (3) & (4) & (5) & (6) & (7) & (8) & (9) & (10) & (11) & (12) & (13) & (14)\\
\hline
   95A &  16.48 &  16.47 &  15.90 &  16.51 &  16.36 &  16.47 &  15.26 &  16.38 &  16.47 &  16.39 &  16.55 &  15.81 & 16.53 \\        &   0.09 &   0.09 &   0.06 &   0.11 &   0.12 &   0.13 &   0.04
&   0.12 &   0.13 &   0.20 &   0.34 &   0.11 &  0.31 \\
    96 &  16.54 &  16.55 &  16.52 &  16.48 &  16.79 &  16.85 &  16.57 &  16.76 &  16.90 &  16.99 &  16.78 &  16.74 & 16.81 \\        &   0.10 &   0.10 &   0.11 &   0.11 &   0.17 &   0.18 &   0.14
&   0.17 &   0.20 &   0.31 &   0.36 &   0.24 &  0.35 \\
    97 &  16.98 &  16.96 &  16.94 &  17.20 &  17.11 &  17.31 &  16.78 &  17.45 &  17.70 &  17.70 &  17.46 &  17.38 & 18.09 \\        &   0.11 &   0.11 &   0.12 &   0.15 &   0.17 &   0.21 &   0.13
&   0.23 &   0.30 &   0.45 &   0.52 &   0.32 &  0.86 \\
    99 &  15.53 &  15.55 &  15.21 &  15.54 &  15.59 &  15.64 &  14.71 &  15.60 &  15.55 &  15.90 &  16.24 &  15.60 & 16.08 \\        &   0.04 &   0.05 &   0.04 &   0.05 &   0.06 &   0.07 &   0.03
&   0.07 &   0.07 &   0.12 &   0.23 &   0.09 &  0.19 \\
   100 &  14.49 &  14.46 &  14.25 &  14.49 &  14.47 &  14.55 &  13.66 &  14.53 &  14.55 &  14.51 &  14.56 &  14.25 & 14.43 \\        &   0.04 &   0.04 &   0.03 &   0.04 &   0.05 &   0.05 &   0.02
&   0.05 &   0.05 &   0.08 &   0.12 &   0.06 &  0.10 \\      
   200 &  16.32 &  16.35 &  16.01 &  16.44 &  16.54 &  16.64 &  15.41 &  16.82 &  17.06 &  17.12 &  17.26 &  16.40 & 17.05 \\        &   0.07 &   0.08 &   0.06 &   0.09 &   0.12 &   0.13 &   0.04
&   0.15 &   0.19 &   0.32 &   0.53 &   0.16 &  0.41 \\
   201 &  16.91 &  17.10 &  16.67 &  17.16 &  17.09 &  17.40 &  16.44 &  17.34 &  17.43 &  17.17 &  17.46 &  16.86 & 16.94 \\        &   0.15 &   0.18 &   0.13 &   0.21 &   0.24 &   0.33 &   0.13
&   0.30 &   0.33 &   0.42 &   0.79 &   0.29 &  0.46 \\
   204 &  14.48 &  14.46 &  14.33 &  14.37 &  14.35 &  14.40 &  14.21 &  14.33 &  14.29 &  14.25 &  14.21 &  14.03 & 14.04 \\        &   0.03 &   0.02 &   0.02 &   0.02 &   0.03 &   0.03 &   0.03
&   0.03 &   0.03 &   0.04 &   0.06 &   0.03 &  0.05 \\
   205 &  14.29 &  14.21 &  14.09 &  14.05 &  14.04 &  14.07 &  14.00 &  14.00 &  13.96 &  13.93 &  13.89 &  13.75 & 13.77 \\        &   0.02 &   0.02 &   0.02 &   0.02 &   0.02 &   0.02 &   0.02
&   0.02 &   0.02 &   0.03 &   0.05 &   0.03 &  0.04 \\
   209 &  15.72 &  15.71 &  15.56 &  15.77 &  15.91 &  16.00 &  15.40 &  15.96 &  15.85 &  15.77 &  15.81 &  15.51 & 15.46 \\        &   0.06 &   0.06 &   0.06 &   0.07 &   0.09 &   0.10 &   0.06
&   0.10 &   0.09 &   0.13 &   0.20 &   0.10 &  0.14 \\
   211 &  17.72 &  17.77 &  17.49 &  17.81 &  17.80 &  17.88 &  16.80 &  17.90 &  18.08 &  18.11 &  18.08 &  17.59 & 18.39 \\        &   0.13 &   0.14 &   0.12 &   0.15 &   0.20 &   0.21 &   0.08
&   0.21 &   0.25 &   0.43 &   0.62 &   0.25 &  0.76 \\
   214 &  16.28 &  16.34 &  15.73 &  16.33 &  16.20 &  16.29 &  14.80 &  16.08 &  16.08 &  16.01 &  15.90 &  15.27 & 15.74 \\        &   0.05 &   0.06 &   0.04 &   0.06 &   0.07 &   0.07 &   0.02
&   0.06 &   0.06 &   0.09 &   0.12 &   0.04 &  0.10 \\
   216 &  16.49 &  16.45 &  15.89 &  16.27 &  16.29 &  16.40 &  15.54 &  16.19 &  16.08 &  16.01 &  15.94 &  15.63 & 15.66 \\        &   0.11 &   0.11 &   0.07 &   0.10 &   0.13 &   0.15 &   0.07
&   0.12 &   0.11 &   0.18 &   0.25 &   0.12 &  0.18 \\       
   217 &  15.57 &  15.56 &  14.80 &  15.53 &  15.51 &  15.60 &  14.64 &  15.54 &  15.53 &  15.51 &  15.56 &  15.15 & 15.49 \\        &   0.06 &   0.06 &   0.03 &   0.06 &   0.08 &   0.08 &   0.03
&   0.07 &   0.07 &   0.13 &   0.20 &   0.08 &  0.17 \\
   218 &  14.66 &  14.68 &  14.19 &  14.69 &  14.77 &  14.84 &  13.78 &  14.81 &  14.82 &  14.77 &  14.80 &  14.41 & 14.73 \\        &   0.03 &   0.03 &   0.02 &   0.03 &   0.05 &   0.05 &   0.02
&   0.04 &   0.05 &   0.08 &   0.12 &   0.05 &  0.10 \\
   220 &  16.58 &  16.58 &  15.85 &  16.55 &  16.51 &  16.59 &  15.55 &  16.40 &  16.35 &  16.25 &  16.31 &  15.71 & 16.05 \\        &   0.07 &   0.07 &   0.04 &   0.07 &   0.09 &   0.10 &   0.04
&   0.08 &   0.08 &   0.12 &   0.19 &   0.07 &  0.14 \\
   221 &  16.03 &  16.05 &  15.47 &  15.96 &  15.97 &  16.00 &  15.03 &  15.73 &  15.53 &  15.40 &  15.39 &  14.91 & 14.98 \\        &   0.06 &   0.06 &   0.04 &   0.06 &   0.08 &   0.08 &   0.03
&   0.06 &   0.05 &   0.08 &   0.12 &   0.05 &  0.07 \\
   222 &  17.01 &  17.05 &  16.10 &  17.25 &  17.06 &  17.35 &  15.38 &  17.15 &  17.19 &  17.04 &  16.99 &  16.04 & 16.81 \\        &   0.15 &   0.16 &   0.07 &   0.20 &   0.23 &   0.31 &   0.05
&   0.24 &   0.25 &   0.40 &   0.57 &   0.15 &  0.44 \\
\end{tabular}
\end{table}                  
}
 
\clearpage
{\small 
\setcounter{table}{2}
\begin{table}[ht]
\caption{Continued}
\vspace {0.3cm}
\begin{tabular}{cccccccccccccc}
\hline
\hline
 No. & 03  &  04 &  05 &  06 &  07 &  08 &  09 &  10 &  11 &  12 &  13 &
14 &  15\\
(1)    & (2) & (3) & (4) & (5) & (6) & (7) & (8) & (9) & (10) & (11) & (12) & (13) & (14)\\
\hline
   224 &  17.69 &  17.62 &  17.43 &  17.64 &  17.52 &  17.69 &  16.78 &  17.30 &  17.10 &  16.97 &  16.84 &  16.40 & 16.45 \\        &   0.21 &   0.20 &   0.18 &   0.22 &   0.26 &   0.31 &   0.13
&   0.21 &   0.18 &   0.27 &   0.36 &   0.15 &  0.23 \\
   226 &  17.41 &  17.37 &  17.12 &  17.33 &  17.18 &  17.12 &  16.34 &  16.65 &  16.43 &  16.31 &  16.22 &  16.08 & 15.99 \\        &   0.15 &   0.15 &   0.13 &   0.15 &   0.18 &   0.17 &   0.08
&   0.11 &   0.09 &   0.14 &   0.19 &   0.10 &  0.14 \\
   230 &  15.61 &  15.56 &  15.26 &  15.41 &  15.28 &  15.35 &  14.91 &  15.25 &  15.23 &  15.10 &  15.18 &  14.94 & 15.03 \\        &   0.06 &   0.06 &   0.05 &   0.06 &   0.07 &   0.07 &   0.05
&   0.06 &   0.06 &   0.10 &   0.16 &   0.08 &  0.12 \\
   231 &  17.98 &  17.96 &  17.63 &  18.09 &  17.97 &  18.10 &  17.02 &  17.92 &  18.02 &  17.90 &  17.88 &  17.56 & 17.99 \\        &   0.19 &   0.19 &   0.15 &   0.22 &   0.27 &   0.31 &   0.11
&   0.25 &   0.27 &   0.44 &   0.65 &   0.30 &  0.65 \\
   232 &  17.18 &  17.16 &  16.98 &  17.15 &  16.98 &  17.01 &  16.59 &  16.76 &  16.75 &  16.66 &  16.58 &  16.45 & 16.43 \\        &   0.11 &   0.10 &   0.09 &   0.11 &   0.12 &   0.13 &   0.08
&   0.09 &   0.10 &   0.15 &   0.21 &   0.12 &  0.17 \\     
   233 &  17.32 &  17.24 &  16.71 &  17.14 &  16.81 &  16.87 &  15.72 &  16.28 &  15.86 &  15.74 &  15.78 &  15.47 & 15.42 \\        &   0.24 &   0.22 &   0.15 &   0.21 &   0.22 &   0.23 &   0.08
&   0.13 &   0.09 &   0.14 &   0.22 &   0.10 &  0.15 \\
   234 &  16.09 &  16.13 &  16.07 &  16.24 &  16.29 &  16.39 &  16.15 &  16.50 &  16.59 &  16.57 &  16.70 &  16.59 & 16.67 \\        &   0.05 &   0.05 &   0.05 &   0.06 &   0.08 &   0.09 &   0.07
&   0.09 &   0.10 &   0.17 &   0.28 &   0.16 &  0.25 \\
   236 &  16.66 &  16.69 &  16.67 &  16.90 &  16.90 &  17.02 &  16.63 &  17.12 &  17.25 &  17.17 &  17.41 &  17.13 & 17.90 \\        &   0.09 &   0.09 &   0.10 &   0.12 &   0.16 &   0.18 &   0.12
&   0.18 &   0.21 &   0.35 &   0.66 &   0.31 &  0.95 \\
   237 &  16.01 &  16.06 &  15.62 &  16.11 &  16.06 &  16.15 &  15.49 &  16.05 &  16.03 &  15.97 &  16.00 &  15.69 & 15.81 \\        &   0.06 &   0.07 &   0.05 &   0.07 &   0.09 &   0.10 &   0.05
&   0.09 &   0.09 &   0.15 &   0.23 &   0.11 &  0.18 \\
   238 &  16.20 &  16.21 &  15.94 &  16.28 &  16.27 &  16.43 &  15.80 &  16.51 &  16.63 &  16.64 &  16.73 &  16.61 & 16.83 \\        &   0.05 &   0.05 &   0.04 &   0.05 &   0.07 &   0.08 &   0.04
&   0.08 &   0.09 &   0.15 &   0.25 &   0.14 &  0.25 \\
   240 &  17.83 &  17.86 &  16.95 &  17.98 &  17.85 &  18.04 &  16.71 &  17.93 &  18.02 &  17.95 &  18.02 &  17.34 & 17.92 \\        &   0.15 &   0.15 &   0.08 &   0.18 &   0.22 &   0.26 &   0.08
&   0.22 &   0.24 &   0.40 &   0.65 &   0.21 &  0.54 \\
   241 &  18.07 &  18.38 &  17.62 &  18.28 &  18.20 &  18.15 &  17.18 &  18.32 &  18.48 &  18.17 &  18.46 &  18.10 & 18.20 \\        &   0.17 &   0.21 &   0.12 &   0.21 &   0.26 &   0.25 &   0.10
&   0.27 &   0.32 &   0.44 &   0.87 &   0.38 &  0.62 \\
   242 &  16.66 &  16.64 &  16.58 &  16.64 &  16.72 &  16.82 &  16.77 &  16.80 &  16.76 &  16.73 &  16.79 &  16.54 & 16.49 \\        &   0.09 &   0.09 &   0.09 &   0.09 &   0.13 &   0.15 &   0.13
&   0.14 &   0.13 &   0.23 &   0.36 &   0.18 &  0.25 \\     
   243 &  16.50 &  16.54 &  16.22 &  16.60 &  16.66 &  16.78 &  16.36 &  16.91 &  17.11 &  17.17 &  17.26 &  16.98 & 17.35 \\        &   0.08 &   0.08 &   0.06 &   0.09 &   0.12 &   0.14 &   0.09
&   0.15 &   0.18 &   0.34 &   0.57 &   0.26 &  0.56 \\
   245 &  16.61 &  16.42 &  16.19 &  16.46 &  16.26 &  16.35 &  15.77 &  16.16 &  16.19 &  16.10 &  16.13 &  15.83 & 16.02 \\        &   0.19 &   0.16 &   0.14 &   0.18 &   0.20 &   0.22 &   0.12
&   0.17 &   0.18 &   0.30 &   0.46 &   0.22 &  0.39 \\
   246 &  15.75 &  15.72 &  15.58 &  16.04 &  16.10 &  16.05 &  15.69 &  16.39 &  16.57 &  16.54 &  16.89 &  16.60 & 17.03 \\        &   0.05 &   0.05 &   0.05 &   0.07 &   0.09 &   0.09 &   0.06
&   0.11 &   0.13 &   0.23 &   0.48 &   0.22 &  0.50 \\
   255 &  16.46 &  16.51 &  15.55 &  16.41 &  16.32 &  16.41 &  15.15 &  16.14 &  16.15 &  16.03 &  16.05 &  15.56 & 15.96 \\        &   0.09 &   0.10 &   0.05 &   0.09 &   0.12 &   0.13 &   0.04
&   0.09 &   0.10 &   0.16 &   0.24 &   0.09 &  0.21 \\
   256 &  15.12 &  15.03 &  14.62 &  14.94 &  14.88 &  14.91 &  14.33 &  14.84 &  14.94 &  14.85 &  14.90 &  14.74 & 14.97 \\        &   0.03 &   0.03 &   0.02 &   0.03 &   0.04 &   0.04 &   0.02
&   0.03 &   0.04 &   0.06 &   0.10 &   0.05 &  0.09 \\
\end{tabular}
\end{table}            
}
 
\clearpage
{\small  
\setcounter{table}{2}
\begin{table}[ht]
\caption{Continued}
\vspace {0.3cm}
\begin{tabular}{cccccccccccccc}
\hline
\hline
 No. & 03  &  04 &  05 &  06 &  07 &  08 &  09 &  10 &  11 &  12 &  13 &
14 &  15\\
(1)    & (2) & (3) & (4) & (5) & (6) & (7) & (8) & (9) & (10) & (11) & (12) & (13) & (14)\\
\hline
   258 &  17.52 &  17.21 &  16.28 &  16.97 &  16.57 &  16.48 &  15.74 &  16.17 &  16.09 &  15.97 &  15.94 &  15.66 & 15.86 \\        &   0.20 &   0.15 &   0.07 &   0.13 &   0.13 &   0.12 &   0.06
&   0.08 &   0.08 &   0.13 &   0.19 &   0.09 &  0.16 \\
   263 &  16.13 &  16.08 &  15.87 &  16.00 &  15.87 &  15.87 &  15.15 &  15.54 &  15.33 &  15.24 &  15.23 &  14.90 & 14.93 \\        &   0.11 &   0.10 &   0.09 &   0.10 &   0.12 &   0.13 &   0.06
&   0.09 &   0.07 &   0.13 &   0.19 &   0.08 &  0.13 \\
   264 &  16.13 &  16.14 &  15.81 &  16.16 &  16.23 &  16.30 &  15.70 &  16.41 &  16.52 &  16.46 &  16.54 &  16.41 & 16.54 \\        &   0.06 &   0.06 &   0.05 &   0.07 &   0.10 &   0.11 &   0.06
&   0.11 &   0.12 &   0.21 &   0.35 &   0.18 &  0.32 \\
   268 &  15.03 &  14.98 &  14.85 &  14.95 &  14.96 &  15.01 &  14.96 &  15.06 &  15.12 &  15.11 &  15.11 &  15.07 & 15.09 \\        &   0.02 &   0.02 &   0.02 &   0.02 &   0.03 &   0.03 &   0.03
&   0.03 &   0.03 &   0.05 &   0.08 &   0.05 &  0.07 \\
   272 &  16.09 &  16.02 &  15.43 &  15.94 &  15.83 &  15.83 &  15.04 &  15.74 &  15.75 &  15.64 &  15.67 &  15.35 & 15.65 \\        &   0.07 &   0.07 &   0.05 &   0.07 &   0.09 &   0.09 &   0.04
&   0.07 &   0.07 &   0.13 &   0.20 &   0.09 &  0.18 \\          
   274 &  14.43 &  14.37 &  13.80 &  14.31 &  14.32 &  14.37 &  13.65 &  14.27 &  14.29 &  14.21 &  14.26 &  14.03 & 14.32 \\        &   0.04 &   0.04 &   0.02 &   0.04 &   0.05 &   0.05 &   0.03
&   0.04 &   0.05 &   0.08 &   0.13 &   0.06 &  0.13 \\
   275 &  17.77 &  17.72 &  17.44 &  17.76 &  17.81 &  17.81 &  17.22 &  17.78 &  17.71 &  17.85 &  17.70 &  17.72 & 17.29 \\        &   0.16 &   0.16 &   0.13 &   0.17 &   0.25 &   0.25 &   0.13
&   0.22 &   0.21 &   0.46 &   0.61 &   0.37 &  0.38 \\
   276 &  16.56 &  16.29 &  16.03 &  16.11 &  15.87 &  15.98 &  15.91 &  15.99 &  16.02 &  15.86 &  15.90 &  15.79 & 16.18 \\        &   0.16 &   0.12 &   0.10 &   0.11 &   0.12 &   0.14 &   0.12
&   0.13 &   0.13 &   0.21 &   0.34 &   0.18 &  0.40 \\
   277 &  13.59 &  13.46 &  13.09 &  13.30 &  13.27 &  13.28 &  12.85 &  13.21 &  13.24 &  13.19 &  13.20 &  13.07 & 13.17 \\        &   0.01 &   0.01 &   0.01 &   0.01 &   0.02 &   0.02 &   0.01
&   0.01 &   0.01 &   0.02 &   0.03 &   0.02 &  0.03 \\
   278 &  18.49 &  18.36 &  17.97 &  18.13 &  17.80 &  17.83 &  17.42 &  17.85 &  17.91 &  17.75 &  17.83 &  17.62 & 17.90 \\        &   0.26 &   0.23 &   0.18 &   0.20 &   0.20 &   0.21 &   0.14
&   0.20 &   0.22 &   0.32 &   0.51 &   0.26 &  0.50 \\
   280 &  14.86 &  14.91 &  13.24 &  15.00 &  14.97 &  15.17 &  13.54 &  15.10 &  15.26 &  15.22 &  15.44 &  14.50 & 15.47 \\        &   0.03 &   0.03 &   0.01 &   0.03 &   0.04 &   0.05 &   0.01
&   0.05 &   0.05 &   0.09 &   0.18 &   0.05 &  0.16 \\
   285 &  19.25 &  19.06 &  18.74 &  18.84 &  18.54 &  18.68 &  18.49 &  18.47 &  18.40 &  18.22 &  18.14 &  18.23 & 18.42 \\        &   0.53 &   0.44 &   0.36 &   0.38 &   0.41 &   0.47 &   0.36
&   0.36 &   0.35 &   0.54 &   0.76 &   0.50 &  0.90 \\
   288 &  15.56 &  15.59 &  15.11 &  15.66 &  15.70 &  15.79 &  15.04 &  15.74 &  15.68 &  15.63 &  15.60 &  15.35 & 15.48 \\        &   0.06 &   0.06 &   0.04 &   0.06 &   0.09 &   0.10 &   0.05
&   0.09 &   0.09 &   0.15 &   0.22 &   0.11 &  0.18 \\
  290A &  16.10 &  16.08 &  15.06 &  16.26 &  16.16 &  16.45 &  15.26 &  16.51 &  16.65 &  16.64 &  16.75 &  16.27 & 17.06 \\        &   0.08 &   0.08 &   0.04 &   0.09 &   0.12 &   0.16 &   0.05
&   0.15 &   0.18 &   0.32 &   0.53 &   0.21 &  0.65 \\   
  290B &  17.81 &  17.81 &  17.75 &  17.92 &  17.91 &  18.06 &  17.50 &  18.17 &  18.31 &  18.24 &  18.42 &  18.39 & 18.55 \\        &   0.13 &   0.13 &   0.14 &   0.15 &   0.20 &   0.24 &   0.14
&   0.24 &   0.28 &   0.47 &   0.83 &   0.49 &  0.86 \\
   301 &  14.16 &  14.07 &  13.79 &  13.95 &  13.94 &  13.95 &  13.18 &  13.90 &  13.94 &  13.91 &  13.85 &  13.67 & 13.90 \\        &   0.02 &   0.02 &   0.02 &   0.02 &   0.02 &   0.02 &   0.01
&   0.02 &   0.02 &   0.03 &   0.04 &   0.02 &  0.04 \\
   302 &  15.35 &  15.41 &  14.56 &  15.58 &  15.45 &  15.70 &  13.82 &  15.52 &  15.63 &  15.41 &  15.57 &  14.80 & 15.44 \\        &   0.05 &   0.06 &   0.03 &   0.07 &   0.08 &   0.10 &   0.02
&   0.08 &   0.10 &   0.13 &   0.22 &   0.07 &  0.18 \\
   601 &  17.60 &  17.50 &  18.02 &  17.50 &  17.37 &  17.38 &  16.82 &  17.25 &  17.24 &  17.16 &  17.17 &  17.04 & 16.75 \\        &   0.22 &   0.20 &   0.36 &   0.21 &   0.26 &   0.27 &   0.15
&   0.22 &   0.22 &   0.38 &   0.58 &   0.31 &  0.36 \\
\end{tabular}
\end{table}            
}
 
\clearpage
{\small 
\setcounter{table}{2}
\begin{table}[ht]
\caption{Continued}
\vspace {0.3cm}
\begin{tabular}{cccccccccccccc}
\hline
\hline
 No. & 03  &  04 &  05 &  06 &  07 &  08 &  09 &  10 &  11 &  12 &  13 &
14 &  15\\
(1)    & (2) & (3) & (4) & (5) & (6) & (7) & (8) & (9) & (10) & (11) & (12) & (13) & (14)\\
\hline
   608 &  16.51 &  16.27 &  15.99 &  15.94 &  15.56 &  15.48 &  14.87 &  14.89 &  14.50 &  14.32 &  14.22 &  14.03 & 13.95 \\        &   0.15 &   0.12 &   0.10 &   0.09 &   0.09 &   0.09 &   0.05
&   0.05 &   0.04 &   0.05 &   0.07 &   0.04 &  0.05 \\
   609 &  15.38 &  15.40 &  15.24 &  15.37 &  15.43 &  15.46 &  15.05 &  15.45 &  15.43 &  15.44 &  15.38 &  15.30 & 15.33 \\        &   0.05 &   0.06 &   0.05 &   0.06 &   0.08 &   0.09 &   0.05
&   0.08 &   0.08 &   0.14 &   0.20 &   0.12 &  0.18 \\
   618 &  17.75 &  17.59 &  17.03 &  17.40 &  17.12 &  17.14 &  16.35 &  17.04 &  17.12 &  16.98 &  17.00 &  16.77 & 16.83 \\        &   0.19 &   0.16 &   0.11 &   0.14 &   0.15 &   0.16 &   0.07
&   0.13 &   0.15 &   0.24 &   0.37 &   0.18 &  0.29 \\
   623 &  15.37 &  15.44 &  14.03 &  15.50 &  15.50 &  15.65 &  14.15 &  15.51 &  15.67 &  15.60 &  15.57 &  14.87 & 15.77 \\        &   0.04 &   0.05 &   0.02 &   0.05 &   0.07 &   0.08 &   0.02
&   0.06 &   0.07 &   0.13 &   0.19 &   0.06 &  0.21 \\
   624 &  16.47 &  16.41 &  15.97 &  16.27 &  16.14 &  16.11 &  15.52 &  15.85 &  15.54 &  15.47 &  15.43 &  15.01 & 15.09 \\        &   0.06 &   0.06 &   0.04 &   0.05 &   0.06 &   0.06 &   0.03
&   0.05 &   0.04 &   0.06 &   0.08 &   0.04 &  0.06 \\    
   625 &  16.71 &  16.68 &  16.35 &  16.61 &  16.57 &  16.53 &  15.97 &  16.29 &  15.98 &  15.90 &  15.84 &  15.45 & 15.52 \\        &   0.07 &   0.07 &   0.06 &   0.07 &   0.09 &   0.09 &   0.05
&   0.07 &   0.05 &   0.09 &   0.12 &   0.05 &  0.08 \\
   626 &  15.34 &  15.31 &  14.93 &  15.26 &  15.24 &  15.33 &  14.69 &  15.30 &  15.31 &  15.33 &  15.44 &  15.20 & 15.37 \\        &   0.05 &   0.05 &   0.04 &   0.05 &   0.07 &   0.08 &   0.04
&   0.07 &   0.07 &   0.13 &   0.22 &   0.11 &  0.19 \\
   629 &  16.34 &  16.35 &  15.90 &  16.44 &  16.40 &  16.52 &  15.75 &  16.52 &  16.55 &  16.43 &  16.55 &  16.11 & 16.43 \\        &   0.05 &   0.05 &   0.04 &   0.06 &   0.08 &   0.09 &   0.04
&   0.08 &   0.08 &   0.13 &   0.21 &   0.09 &  0.18 \\
   631 &  17.26 &  17.26 &  16.44 &  17.37 &  17.29 &  17.43 &  16.15 &  17.35 &  17.42 &  17.35 &  17.47 &  17.03 & 17.48 \\        &   0.14 &   0.14 &   0.08 &   0.17 &   0.21 &   0.24 &   0.07
&   0.21 &   0.23 &   0.38 &   0.62 &   0.26 &  0.58 \\
   632 &  14.99 &  14.96 &  14.81 &  14.94 &  14.98 &  15.05 &  14.64 &  15.11 &  15.17 &  15.18 &  15.27 &  15.15 & 15.38 \\        &   0.04 &   0.04 &   0.03 &   0.04 &   0.05 &   0.06 &   0.04
&   0.06 &   0.06 &   0.10 &   0.17 &   0.09 &  0.17 \\
   635 &  15.09 &  14.89 &  14.75 &  14.68 &  14.67 &  14.65 &  14.57 &  14.63 &  14.60 &  14.58 &  14.57 &  14.60 & 14.57 \\        &   0.04 &   0.03 &   0.03 &   0.03 &   0.03 &   0.03 &   0.03
&   0.03 &   0.03 &   0.05 &   0.08 &   0.05 &  0.07 \\
   637 &  16.35 &  16.07 &  14.87 &  15.96 &  15.67 &  15.65 &  14.89 &  15.55 &  15.50 &  15.41 &  15.42 &  15.19 & 15.54 \\        &   0.14 &   0.10 &   0.04 &   0.10 &   0.11 &   0.11 &   0.05
&   0.09 &   0.09 &   0.15 &   0.23 &   0.11 &  0.24 \\
   638 &  15.35 &  15.39 &  13.24 &  15.23 &  15.26 &  15.41 &  13.67 &  15.34 &  15.44 &  15.30 &  15.43 &  14.42 & 15.45 \\        &   0.05 &   0.06 &   0.01 &   0.05 &   0.07 &   0.09 &   0.02
&   0.07 &   0.08 &   0.13 &   0.23 &   0.05 &  0.21 \\       
   641 &  15.63 &  15.66 &  14.87 &  15.83 &  15.77 &  15.93 &  14.79 &  16.04 &  16.07 &  16.04 &  16.27 &  15.63 & 16.30 \\        &   0.09 &   0.10 &   0.05 &   0.12 &   0.16 &   0.19 &   0.06
&   0.19 &   0.20 &   0.37 &   0.70 &   0.23 &  0.67 \\
   648 &  16.94 &  16.96 &  16.89 &  16.99 &  16.98 &  16.95 &  16.44 &  16.94 &  16.77 &  16.62 &  17.08 &  16.69 & 16.60 \\        &   0.19 &   0.19 &   0.19 &   0.20 &   0.29 &   0.29 &   0.16
&   0.26 &   0.23 &   0.37 &   0.86 &   0.36 &  0.51 \\
   650 &  15.58 &  15.72 &  14.83 &  15.89 &  15.86 &  16.16 &  14.09 &  16.15 &  16.26 &  16.14 &  16.21 &  15.78 & 16.52 \\        &   0.08 &   0.09 &   0.04 &   0.11 &   0.16 &   0.21 &   0.03
&   0.19 &   0.21 &   0.36 &   0.58 &   0.23 &  0.72 \\
   660 &  16.50 &  16.32 &  15.83 &  16.51 &  16.36 &  16.38 &  15.89 &  16.53 &  16.17 &  16.19 &  15.74 &  15.76 & 15.38 \\        &   0.26 &   0.22 &   0.16 &   0.28 &   0.35 &   0.36 &   0.21
&   0.38 &   0.28 &   0.52 &   0.51 &   0.32 &  0.34 \\
   663 &  17.36 &  17.20 &  16.65 &  16.90 &  16.32 &  16.53 &  15.96 &  16.05 &  15.84 &  15.70 &  15.77 &  15.31 & 15.76 \\        &   0.49 &   0.43 &   0.28 &   0.35 &   0.29 &   0.35 &   0.19
&   0.21 &   0.18 &   0.28 &   0.45 &   0.18 &  0.41 \\
\end{tabular}
\end{table}                           
}
 
\clearpage
{\small 
\setcounter{table}{2}
\begin{table}[ht]
\caption{Continued}
\vspace {0.3cm}
\begin{tabular}{cccccccccccccc}
\hline
\hline
 No. & 03  &  04 &  05 &  06 &  07 &  08 &  09 &  10 &  11 &  12 &  13 &
14 &  15\\
(1)    & (2) & (3) & (4) & (5) & (6) & (7) & (8) & (9) & (10) & (11) & (12) & (13) & (14)\\
\hline
   664 &  15.72 &  15.75 &  15.39 &  15.83 &  15.92 &  16.08 &  15.37 &  16.19 &  16.14 &  16.22 &  15.85 &  15.93 & 16.13 \\        &   0.08 &   0.08 &   0.06 &   0.09 &   0.14 &   0.16 &   0.08
&   0.17 &   0.16 &   0.32 &   0.34 &   0.22 &  0.41 \\
   665 &  15.58 &  15.53 &  15.50 &  15.49 &  15.46 &  15.55 &  15.44 &  15.56 &  15.46 &  15.37 &  15.40 &  15.15 & 15.13 \\        &   0.09 &   0.08 &   0.09 &   0.08 &   0.11 &   0.12 &   0.10
&   0.12 &   0.11 &   0.17 &   0.26 &   0.13 &  0.19 \\
   667 &  15.59 &  15.59 &  15.44 &  15.67 &  15.72 &  15.82 &  15.14 &  15.83 &  15.90 &  15.85 &  15.96 &  15.61 & 15.82 \\        &   0.06 &   0.06 &   0.05 &   0.07 &   0.09 &   0.10 &   0.05
&   0.10 &   0.10 &   0.17 &   0.28 &   0.13 &  0.23 \\
   680 &  12.32 &  12.63 &  11.26 &  12.39 &  12.39 &  12.62 &  10.76 &  12.52 &  12.61 &  12.53 &  12.69 &  11.82 & 12.71 \\        &   0.01 &   0.01 &   0.00 &   0.01 &   0.01 &   0.02 &   0.00
&   0.02 &   0.02 &   0.03 &   0.05 &   0.01 &  0.04 \\
   684 &  16.99 &  17.06 &  16.90 &  17.79 &  16.74 &  16.87 &  15.56 &  16.83 &  16.77 &  16.44 &  16.68 &  16.37 & 16.48 \\        &   0.21 &   0.23 &   0.21 &   0.48 &   0.25 &   0.29 &   0.08
&   0.26 &   0.25 &   0.33 &   0.61 &   0.28 &  0.47 \\      
   685 &  17.06 &  17.22 &  16.83 &  17.63 &  17.32 &  17.59 &  15.87 &  17.55 &  17.57 &  17.31 &  17.51 &  16.86 & 17.37 \\        &   0.14 &   0.16 &   0.13 &   0.26 &   0.26 &   0.34 &   0.07
&   0.31 &   0.32 &   0.45 &   0.80 &   0.27 &  0.65 \\
   687 &  16.42 &  16.07 &  15.74 &  15.83 &  15.49 &  15.45 &  14.93 &  15.24 &  15.16 &  15.07 &  15.01 &  14.87 & 14.90 \\        &   0.11 &   0.08 &   0.07 &   0.07 &   0.07 &   0.07 &   0.04
&   0.05 &   0.05 &   0.08 &   0.11 &   0.06 &  0.09 \\
   688 &  14.72 &  14.69 &  14.60 &  14.66 &  14.69 &  14.77 &  14.67 &  14.83 &  14.89 &  14.86 &  14.85 &  14.79 & 14.91 \\        &   0.02 &   0.02 &   0.02 &   0.03 &   0.03 &   0.03 &   0.03
&   0.03 &   0.04 &   0.06 &   0.08 &   0.05 &  0.08 \\
   689 &  15.50 &  15.51 &  15.46 &  15.55 &  15.59 &  15.70 &  15.55 &  15.61 &  15.52 &  15.54 &  15.53 &  15.46 & 15.38 \\        &   0.04 &   0.04 &   0.05 &   0.05 &   0.06 &   0.07 &   0.06
&   0.06 &   0.06 &   0.09 &   0.13 &   0.08 &  0.11 \\
   692 &  17.53 &  17.32 &  16.96 &  17.14 &  16.84 &  16.90 &  16.35 &  16.73 &  16.62 &  16.42 &  16.37 &  16.25 & 16.41 \\        &   0.24 &   0.20 &   0.16 &   0.18 &   0.18 &   0.20 &   0.11
&   0.16 &   0.15 &   0.21 &   0.29 &   0.17 &  0.28 \\
   696 &  16.92 &  16.90 &  16.76 &  16.96 &  16.98 &  17.05 &  16.25 &  17.09 &  17.08 &  17.11 &  17.11 &  16.79 & 16.91 \\        &   0.11 &   0.11 &   0.10 &   0.12 &   0.16 &   0.18 &   0.08
&   0.17 &   0.18 &   0.30 &   0.45 &   0.21 &  0.34 \\
   704 &  15.17 &  15.18 &  15.08 &  15.13 &  15.20 &  15.32 &  15.31 &  15.47 &  15.61 &  15.73 &  15.89 &  15.87 & 15.77 \\        &   0.06 &   0.06 &   0.06 &   0.07 &   0.09 &   0.11 &   0.10
&   0.12 &   0.14 &   0.25 &   0.43 &   0.26 &  0.35 \\
   707 &  16.29 &  16.16 &  15.89 &  16.09 &  16.14 &  16.29 &  15.66 &  16.20 &  16.17 &  16.29 &  16.33 &  16.22 & 15.83 \\        &   0.14 &   0.12 &   0.11 &   0.13 &   0.17 &   0.20 &   0.11
&   0.18 &   0.18 &   0.31 &   0.46 &   0.27 &  0.27 \\
   709 &  15.40 &  15.33 &  15.22 &  15.28 &  15.34 &  15.42 &  15.05 &  15.51 &  15.56 &  15.43 &  15.69 &  15.64 & 16.48 \\        &   0.06 &   0.05 &   0.05 &   0.05 &   0.07 &   0.08 &   0.06
&   0.08 &   0.09 &   0.13 &   0.25 &   0.15 &  0.47 \\    
   710 &  16.83 &  16.87 &  16.07 &  17.05 &  16.84 &  17.04 &  15.38 &  16.86 &  16.93 &  16.70 &  16.65 &  15.88 & 16.67 \\        &   0.08 &   0.09 &   0.05 &   0.11 &   0.12 &   0.14 &   0.03
&   0.11 &   0.12 &   0.16 &   0.23 &   0.07 &  0.22 \\
   711 &  15.14 &  15.17 &  14.79 &  15.21 &  15.22 &  15.38 &  14.72 &  15.36 &  15.32 &  15.17 &  15.38 &  14.92 & 15.29 \\        &   0.04 &   0.05 &   0.04 &   0.05 &   0.07 &   0.08 &   0.04
&   0.07 &   0.07 &   0.10 &   0.19 &   0.08 &  0.16 \\
   713 &  16.39 &  16.25 &  16.09 &  16.12 &  15.93 &  15.99 &  15.74 &  15.97 &  16.04 &  15.91 &  15.89 &  15.86 & 16.36 \\        &   0.10 &   0.09 &   0.08 &   0.08 &   0.09 &   0.10 &   0.07
&   0.09 &   0.10 &   0.15 &   0.21 &   0.13 &  0.30 \\
   714 &  15.66 &  15.67 &  15.18 &  15.61 &  15.59 &  15.74 &  14.75 &  15.70 &  15.65 &  15.49 &  15.58 &  15.11 & 15.87 \\        &   0.08 &   0.08 &   0.05 &   0.08 &   0.10 &   0.12 &   0.05
&   0.11 &   0.11 &   0.15 &   0.25 &   0.10 &  0.30 \\
\end{tabular}
\end{table}          
}
 
\clearpage
{\small 
\setcounter{table}{2}
\begin{table}[ht]
\caption{Continued}
\vspace {0.3cm}
\begin{tabular}{cccccccccccccc}
\hline
\hline
 No. & 03  &  04 &  05 &  06 &  07 &  08 &  09 &  10 &  11 &  12 &  13 &
14 &  15\\
(1)    & (2) & (3) & (4) & (5) & (6) & (7) & (8) & (9) & (10) & (11) & (12) & (13) & (14)\\
\hline
   720 &  16.74 &  16.65 &  16.32 &  16.74 &  16.62 &  16.76 &  16.01 &  16.73 &  16.84 &  16.74 &  16.75 &  16.72 & 16.67 \\        &   0.09 &   0.09 &   0.07 &   0.10 &   0.12 &   0.14 &   0.07
&   0.13 &   0.14 &   0.21 &   0.32 &   0.20 &  0.27 \\
   728 &  17.46 &  17.54 &  17.38 &  17.69 &  17.47 &  17.74 &  17.19 &  17.65 &  17.25 &  17.23 &  17.35 &  16.95 & 16.97 \\        &   0.21 &   0.22 &   0.21 &   0.27 &   0.31 &   0.40 &   0.22
&   0.34 &   0.24 &   0.44 &   0.75 &   0.31 &  0.49 \\
   734 &  15.90 &  15.94 &  15.81 &  15.95 &  16.00 &  16.04 &  15.60 &  15.94 &  15.78 &  15.75 &  15.69 &  15.36 & 15.47 \\        &   0.05 &   0.05 &   0.05 &   0.06 &   0.08 &   0.08 &   0.05
&   0.07 &   0.06 &   0.10 &   0.14 &   0.07 &  0.11 \\
   740 &  15.89 &  15.97 &  15.20 &  15.95 &  15.84 &  16.03 &  14.60 &  15.94 &  15.97 &  15.93 &  16.11 &  15.35 & 16.17 \\        &   0.06 &   0.07 &   0.04 &   0.07 &   0.09 &   0.10 &   0.03
&   0.09 &   0.09 &   0.16 &   0.29 &   0.09 &  0.28 \\
   741 &  18.91 &  18.87 &  18.59 &  18.55 &  18.00 &  18.02 &  17.55 &  17.43 &  17.18 &  17.05 &  16.87 &  16.56 & 16.63 \\        &   0.29 &   0.28 &   0.24 &   0.22 &   0.19 &   0.19 &   0.12
&   0.11 &   0.09 &   0.13 &   0.17 &   0.08 &  0.12 \\           
   745 &  18.36 &  18.24 &  17.37 &  18.44 &  18.30 &  18.50 &  16.85 &  18.31 &  18.18 &  18.12 &  17.98 &  17.67 & 17.95 \\        &   0.25 &   0.22 &   0.12 &   0.28 &   0.35 &   0.43 &   0.09
&   0.33 &   0.30 &   0.53 &   0.70 &   0.32 &  0.63 \\
   746 &  17.53 &  17.45 &  17.22 &  17.36 &  17.47 &  17.51 &  17.13 &  17.34 &  17.13 &  17.07 &  17.27 &  16.87 & 16.71 \\        &   0.23 &   0.22 &   0.19 &   0.22 &   0.32 &   0.33 &   0.22
&   0.27 &   0.23 &   0.37 &   0.65 &   0.28 &  0.36 \\
   748 &  17.43 &  16.68 &  16.33 &  16.20 &  15.63 &  15.43 &  15.06 &  15.03 &  14.85 &  14.65 &  14.61 &  14.45 & 14.43 \\        &   0.13 &   0.07 &   0.06 &   0.05 &   0.04 &   0.04 &   0.03
&   0.03 &   0.02 &   0.03 &   0.04 &   0.02 &  0.03 \\
   749 &  15.49 &  15.50 &  15.24 &  15.61 &  15.61 &  15.69 &  14.72 &  15.73 &  16.09 &  15.73 &  15.72 &  15.44 & 15.74 \\        &   0.04 &   0.04 &   0.04 &   0.05 &   0.06 &   0.07 &   0.03
&   0.07 &   0.09 &   0.12 &   0.18 &   0.08 &  0.17 \\
   752 &  17.17 &  17.19 &  16.81 &  17.20 &  17.21 &  17.36 &  16.33 &  17.18 &  17.06 &  17.02 &  17.19 &  17.23 & 17.05 \\        &   0.14 &   0.14 &   0.11 &   0.14 &   0.20 &   0.24 &   0.09
&   0.19 &   0.17 &   0.31 &   0.54 &   0.34 &  0.43 \\
   753 &  17.88 &  17.81 &  17.63 &  17.76 &  17.93 &  18.03 &  17.84 &  18.07 &  18.05 &  18.21 &  18.21 &  18.19 & 18.18 \\        &   0.15 &   0.14 &   0.13 &   0.14 &   0.23 &   0.25 &   0.19
&   0.24 &   0.24 &   0.51 &   0.77 &   0.45 &  0.68 \\
   755 &  15.81 &  15.80 &  15.34 &  15.94 &  15.86 &  16.12 &  15.33 &  16.31 &  16.38 &  16.38 &  16.50 &  16.30 & 16.77 \\        &   0.07 &   0.07 &   0.05 &   0.08 &   0.10 &   0.13 &   0.06
&   0.14 &   0.15 &   0.29 &   0.50 &   0.25 &  0.58 \\
   756 &  16.87 &  16.81 &  16.33 &  16.73 &  16.53 &  16.59 &  15.75 &  16.45 &  16.23 &  16.07 &  16.08 &  15.76 & 15.97 \\        &   0.13 &   0.12 &   0.09 &   0.11 &   0.14 &   0.15 &   0.06
&   0.12 &   0.10 &   0.16 &   0.24 &   0.11 &  0.20 \\     
   758 &  16.27 &  16.26 &  15.95 &  16.31 &  16.39 &  16.52 &  15.90 &  16.59 &  16.60 &  16.60 &  16.72 &  16.01 & 16.18 \\        &   0.09 &   0.09 &   0.07 &   0.10 &   0.15 &   0.17 &   0.09
&   0.17 &   0.17 &   0.30 &   0.51 &   0.16 &  0.29 \\
   760 &  16.83 &  16.83 &  16.70 &  16.77 &  16.81 &  16.88 &  16.53 &  16.75 &  16.55 &  16.53 &  16.46 &  16.37 & 16.53 \\        &   0.15 &   0.14 &   0.14 &   0.14 &   0.21 &   0.23 &   0.15
&   0.18 &   0.16 &   0.29 &   0.42 &   0.23 &  0.40 \\
  1002 &  16.76 &  16.64 &  16.41 &  16.65 &  16.45 &  16.59 &  16.31 &  16.54 &  16.48 &  16.42 &  16.38 &  16.14 & 16.20 \\        &   0.11 &   0.10 &   0.09 &   0.11 &   0.11 &   0.13 &   0.10
&   0.12 &   0.12 &   0.18 &   0.24 &   0.13 &  0.19 \\
  1004 &  17.79 &  17.70 &  17.37 &  17.62 &  17.52 &  17.62 &  17.32 &  17.57 &  17.35 &  17.29 &  17.22 &  17.19 & 17.26 \\        &   0.27 &   0.26 &   0.21 &   0.27 &   0.30 &   0.33 &   0.25
&   0.31 &   0.26 &   0.39 &   0.52 &   0.33 &  0.50 \\
  1007 &  16.97 &  16.96 &  16.67 &  17.02 &  17.02 &  17.08 &  15.77 &  16.97 &  16.98 &  16.96 &  17.09 &  16.55 & 16.82 \\        &   0.14 &   0.14 &   0.12 &   0.16 &   0.21 &   0.22 &   0.07
&   0.19 &   0.20 &   0.33 &   0.55 &   0.21 &  0.39 \\
\end{tabular}
\end{table}            
}
 
\clearpage
{\small  
\setcounter{table}{2}
\begin{table}[ht]
\caption{Continued}
\vspace {0.3cm}
\begin{tabular}{cccccccccccccc}
\hline
\hline
 No. & 03  &  04 &  05 &  06 &  07 &  08 &  09 &  10 &  11 &  12 &  13 &
14 &  15\\
(1)    & (2) & (3) & (4) & (5) & (6) & (7) & (8) & (9) & (10) & (11) & (12) & (13) & (14)\\
\hline
  1502 &  15.26 &  15.22 &  15.11 &  15.11 &  15.13 &  15.21 &  15.10 &  15.21 &  15.17 &  15.14 &  15.25 &  15.22 & 15.25 \\        &   0.05 &   0.05 &   0.05 &   0.05 &   0.06 &   0.07 &   0.06
&   0.07 &   0.06 &   0.10 &   0.16 &   0.10 &  0.15 \\
\hline
\end{tabular}
\end{table}
}

\setcounter{figure}{0}
\begin{figure}[ht]
\centerline{\epsfig{file=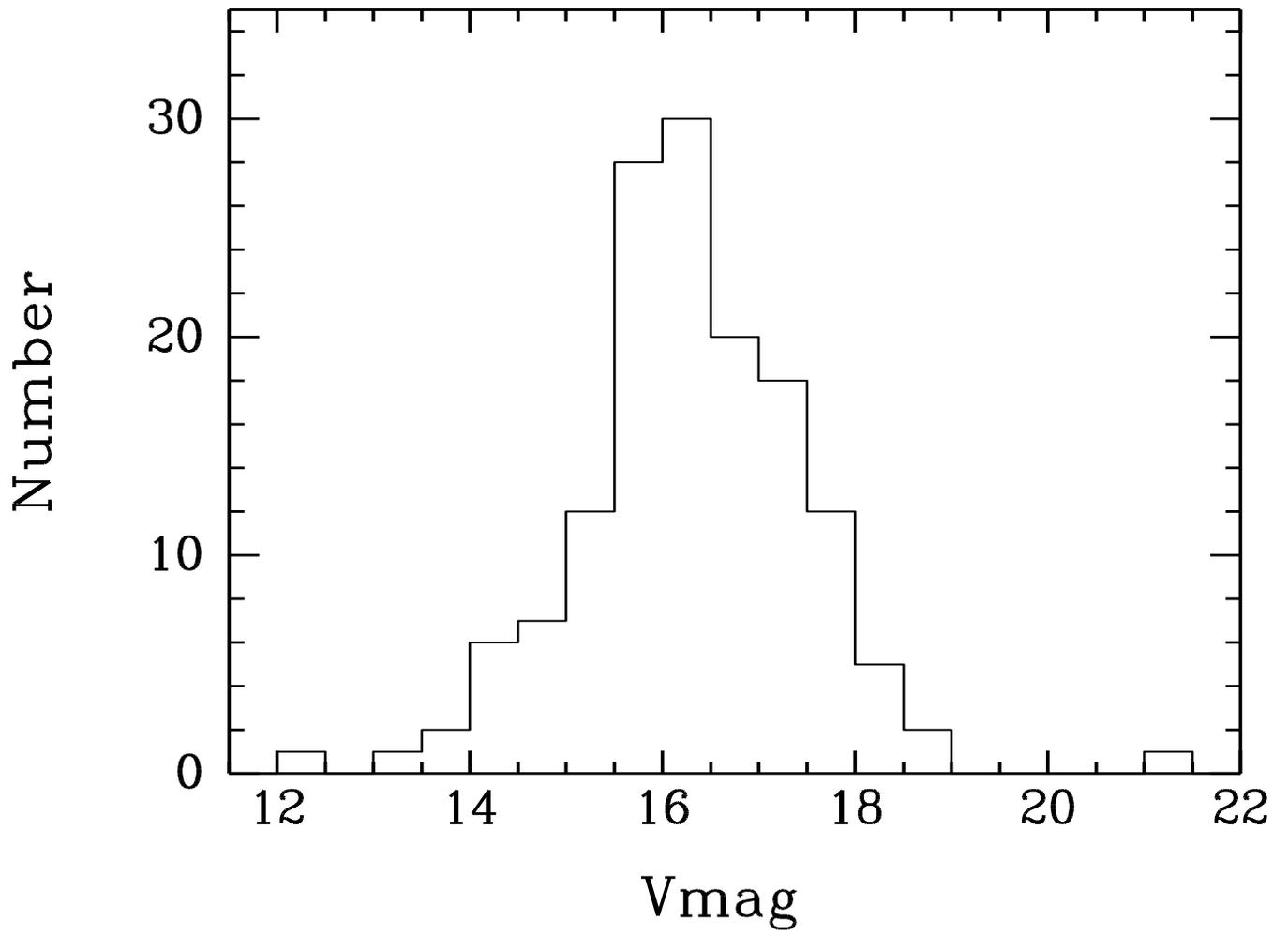,width=16.0cm,bbllx=100,
bblly=134,bburx=510,bbury=668,angle=-90}}
\vspace{-0.5cm}
\caption{Histogram of magnitudes in computed V band
for 145 HII regions.}
\end{figure}
 
\setcounter{figure}{1}
\begin{figure}[ht]
\centerline{\epsfig{file=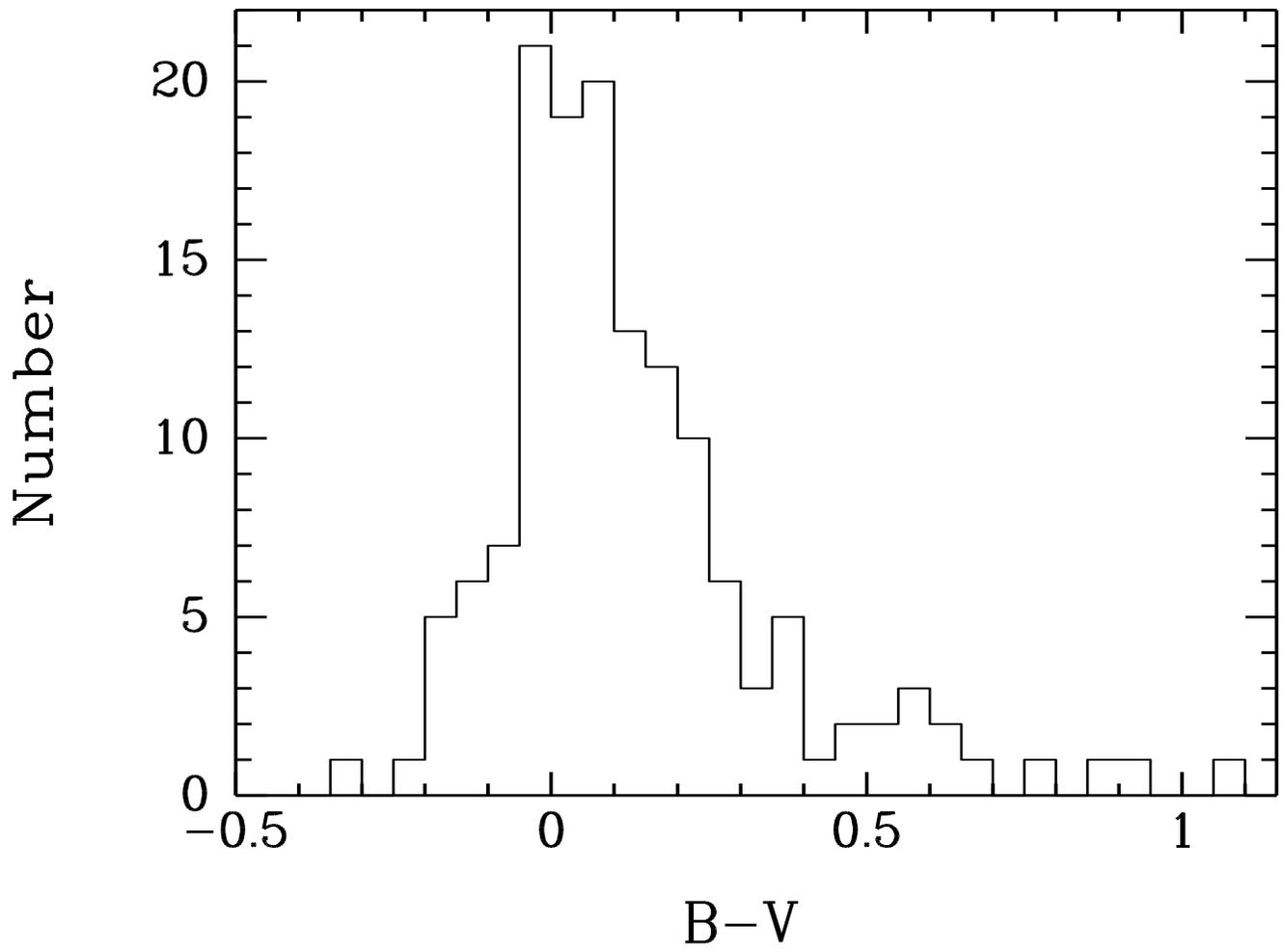,width=16.0cm,bbllx=100,
bblly=134,bburx=510,bbury=668,angle=-90}}
\vspace{-0.5cm}
\caption{Histogram of computed {\it B$-$V} color
for 145 HII regions.}
\end{figure}
 
\setcounter{figure}{2}
\begin{figure}[ht]
\centerline{\epsfig{file=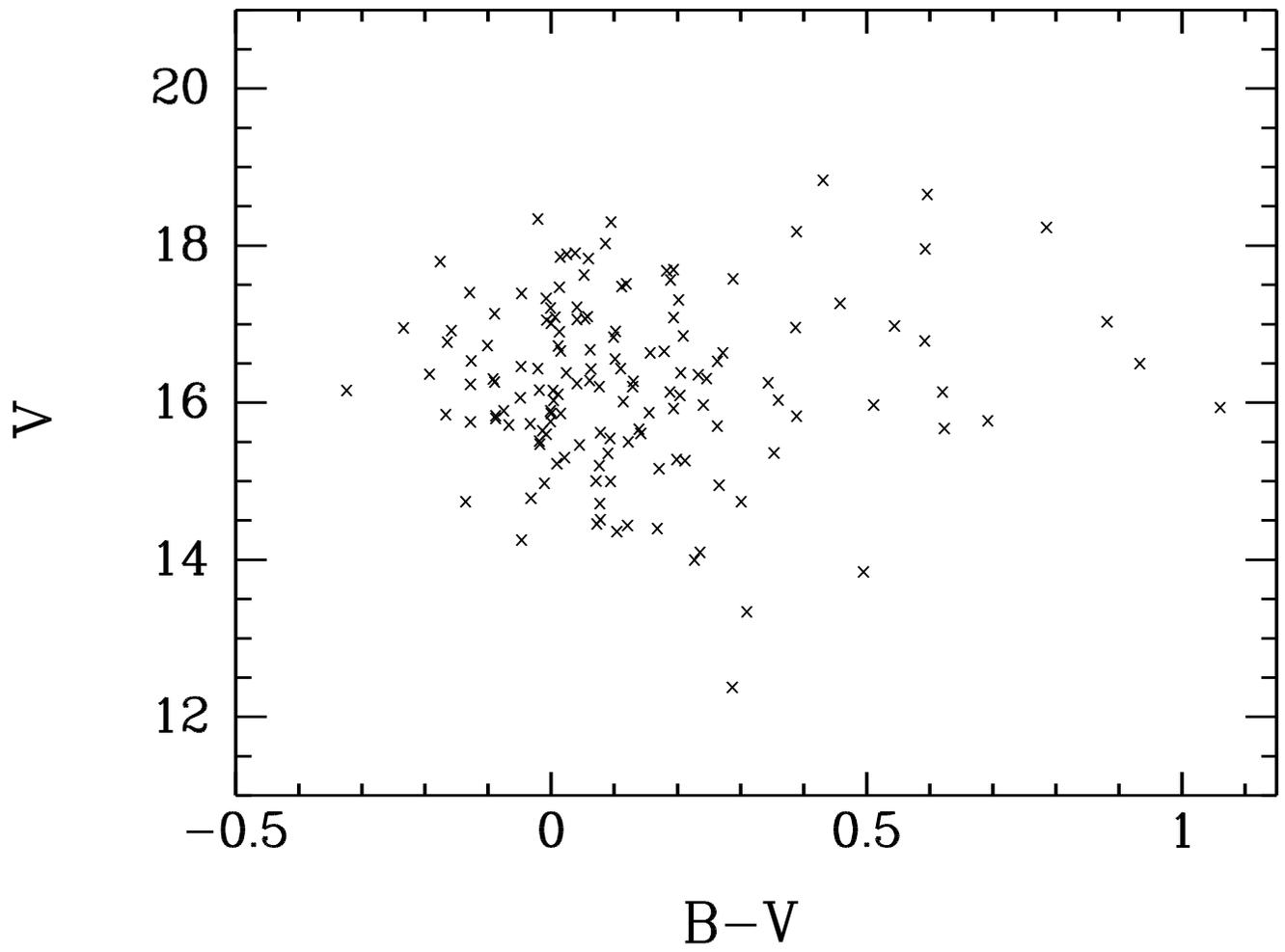,width=16.0cm,bbllx=100,
bblly=134,bburx=510,bbury=668,angle=-90}}
\vspace{-0.5cm}
\caption{The color-magnitude diagram for 145 HII regions.}
\end{figure}

\end{document}